\documentclass[12pt,preprint]{aastex}

\shorttitle{Stochastic Magnetic Field for DSR} \shortauthors{Mao
\& Wang}

\begin{document}

\title{Knot in \object{Cen A}: Stochastic Magnetic Field for
Diffusive Synchrotron Radiation?}

\author{Jirong Mao and Jiancheng Wang}
\affil{Yunnan Observatory, National Astronomical Observatories,
Chinese Academic of Sciences, P.O. Box 110, Kunming, Yunnan
Province, 650011, China}

\email{jirongmao@ynao.ac.cn}

\begin{abstract}
The emission of relativistic electrons moving in the random and
small-scale magnetic field is presented by diffusive synchrotron
radiation (DSR). In this Letter, we revisit the perturbative
treatment of DSR. We propose that random and small-scale magnetic
field might be generated by the turbulence. As an example,
multi-band radiation of the knot in \object{Cen A} comes from the
electrons with energy $\gamma_e\sim 10^3-10^4$ in the magnetic
field of $10^{-3}G$. The multi-band spectrum of DSR is well
determined by the feature of stochastic magnetic field. These
results put strong constraint to the models of particle
acceleration.
\end{abstract}


\section{Introduction}
A popular explanation of non-thermal emission from objects such as
Gamma-Ray Burst (GRB) and jet in Active Galactic Nuclei (AGN) is
synchrotron radiation. The relativistic electrons are buried in
the external, homogenous and steady magnetic field. However, this
large-scale magnetic field is a priori and the origin of it in
GRBs and jets is under debate. Alternatively, the perturbative and
more general non-perturbative treatments of Diffusive Synchrotron
Radiation (DSR) have been proposed
\citep{toptygin87,fleishman06a}, the so-called jitter radiation is
a specific limiting 1D case within the general perturbative DSR
theory. DSR is the emission of relativistic electrons in the local
and random magnetic field. The magnetic field might be produced by
the following process: the anisotropic-distributed plasma can be
disturbed by relativistic collisionless shocks, hence, the initial
magnetic field is produced by the perturbation. The induced
currents from the magnetic field amplify the original magnetic
field thus Weibel instability occurs
\citep{weibel59,medvedev99,frederiksen04,hededal05}. Due to the
lack of the external magnetic field, the particle acceleration can
not be treated by the Feimi-acceleration
\citep{hededal04,nishikawa06} as the usual way. The DSR and jitter
radiation have been selected to predict the spectrum of
GRB/afterglow \citep{fleishman06a,med06,medvedev07,workman07} and
knot in the jet \citep{fleishman06b}. These analytical results are
also identified with numerical simulations \citep{hededal06}.

There are still some problems which should be concentrated on. For
instance, the plasma frequency $\omega_{pe}=(4\pi
e^2n/m_e)^{1/2}$, as a function of the electron density $n$ in the
plasma, is introduced, thus the radiation properties are strongly
affected by the local environment. \citet{medvedev05} played with
the model by merging current filaments to generate the magnetic
field, while in principle the generation of the magnetic field
should be linked with the perturbation of the fluid field. Besides
this filaments merging effect, there could be other possibilities
to produce random magnetic field.

In this paper, following the previous work of \citet{fleishman06a}
and \citet{med06}, we put forward the case of emission by
relativistic electrons moving in stochastic magnetic field. In
Section 2, we review the perturbative DSR and focus on the origin
of the magnetic field. The local and random magnetic field may be
produced by turbulence, but not Weibel instability. In Section 3,
we compare our results to the multi-band spectrum of the knot in
\object{Centaurus A} (hereafter \object{Cen A}). Finally, the
discussion and future expectation are given in Section 4.

\section{Radiation Revisited and Stochastic Magnetic Field}
The emission of single relativistic particle in the small-scale
magnetic field was firstly introduced by \citet{landau71}. Here,
we follow the developed formula to calculate the radiation
intensity, which is the energy per unit frequency per unit time
\citep{fleishman06a}:
\begin{equation}
I_{\omega}
=\frac{e^4}{m^2c^3\gamma^2}\int_{1/2\gamma_\ast^2}^{\infty}d(\frac{\omega'}{\omega})
(\frac{\omega}{\omega'})^2(1-\frac{\omega}{\omega'\gamma_\ast^2}+\frac{\omega^2}{2\omega'^2\gamma_\ast^4})
\times\int
dq_0d\textbf{q}\delta(\omega'-q_0+\textbf{qv})K(\textbf{q})\delta(q_0-q_0(\textbf{q}))
\end{equation}
where
$\omega'=(\omega/2)(\gamma^{-2}+\theta^2+\omega_{pe}^2/\omega^2)$,
$\theta$ is the angle between the electron velocity and the
radiation direction, $\textbf{q}$ and $q_0$ are the wave number
and frequency of the disturbed field respectively,
$\gamma_\ast^{-2}=\gamma^2+\omega_{pe}^2/\omega^2$ and $\gamma$ is
the electron energy, $K(\textbf{q})$ is the term for the random
magnetic field.

Equation (1) is the general expression for perturbative treatment.
It is pointed out by \citet{fleishman06a}, the rectilinear motion
of electron is valid for large frequencies, however, at low
frequencies, the particle trajectory traverses several correlation
lengths scattering by magnetic inhomogeneities, thus the particle
deflection angle accumulated along the coherence length exceeds
the beaming angle (see Figure 1 of \citet{fleishman06a}).

The dispersion relation $q_0=q_0(q)$ of the non-relativistic
plasma was presented in \citet{weibel59}. The improved equations
for the isotropic and relativistic plasma were given in detail by
\citet{mikhailovski80} while \citet{yoon87} built the analytical
model for the relativistic plasma with a waterbag distribution.
More comprehensive works have been performed recently by
\citet{silva02}, \citet{wiersma04} and \citet{fiore06}. In this
paper, we choose the dispersion relation of relativistic
collisionless shocks considered by \citet{milosavlj06}.

Weibel instability is an efficient way to generate the random
magnetic field in relativistic shocks
\citep{silva03,schlickeiser03,wiersma04,lyubarsky06}. However,
there could be other possibilities to form magnetic structure. In
this Letter, we argue that the local and random magnetic field
generated by turbulence is also relevant for the perturbative DSR
theory.

The spectrum energy in a fully developed turbulent fluid can be
described by the Kolmogorov form with the classical index $-5/3$.
For the magnetic turbulence, the cascade delay time may enter the
estimation of energy transfer rate, the energy spectrum of
Kraichnan has an index of $-3/2$. Although the situation we focus
on has no external magnetic field, at small scales, the turbulence
is still shown as the cascade properties. Self-excited
Alfv$\acute{e}$n turbulence has also been found \citep{sokolov06}.
Moreover, we note that the non-magnetized and magnetized
turbulence have a high-degree similarity \citep{cho02,lazarian04}.
All these evidences indicate that a general form of fluid
turbulence can also be valid for the study of random magnetic
field generation. Furthermore, the index of the turbulence
spectrum is not universal. \citet{zhou90} investigated local
turbulent effects with transport models and other nonlinear terms.
Using the scaling model \citep{she94} which presents the cascade
as an infinitely-divisible log-Poisson process \citep{she95},
\citet{boldyrev02} derived a steeper spectrum compared to that of
Kolmogorov. In fact, as estimated by \citet{wang02}, the index
value of a turbulent spectrum has the range between $-1$ to $-2$.
\citet{maclow00} found that the local turbulent spectrum does not
show a straight power-law. Therefore, we propose that the
turbulent spectrum be shown as:
\begin{equation}
F(k)\propto k^{-\alpha}f(k/k_\eta)
\end{equation}
where $k_\nu<k<k_\eta$, $k_\nu$ corresponds to the viscous scale
of the fluid while $k_\eta$ is linked with the scale of the
resistive cascade transfer. We choose $f(k/k_\eta)$ as an
exponential-drop form.

The magnetic field amplified by the turbulence spectrum has been
described by \citet{niemiec04, niemiec06}. We obtain the amplified
magnetic field as:
\begin{equation}
\langle\delta B^2(k)\rangle\propto \int_k^{\infty} F(k')dk'
\end{equation}
In general, Eq.(2) presented as a power-law with a cutoff at high
wave number is universal for the fluid dynamo turbulence, whatever
the radiation field is. The $K(\textbf{q})$ in Eq.(1) can be
linked by the magnetic field as $K(\textbf{q})=C_0\langle\delta
B^2(q)\rangle$ where $C_0$ is the normalization number. Therefore,
this turbulent approach for obtaining magnetic field is the
developments in the framework of current DSR theory.

\section{The Case of Cen A}
\object{Cen A}, the nearest proto-\object{FR}\textrm{I} galaxy,
was sketched from the observational view \citep{israel98}. In
particular, the knot in the jet has been detected in radio, X-ray
\citep{hardcastle03,kraft03,kataoka06} and infrared
\citep{hardcastle06} bands. With these observations, this object
provides a multi-band spectrum to constrain the radiation
mechanisms and the models of particle acceleration.

The central density of the knot in \object{Cen A} is $n=3.7\times
10^{-2}cm^{-3}$ \citep{kraft03}, the correlation length of the
random magnetic field is estimated by
$l_{cor}\sim(0.1-1)l_{sk}\sim 10^5-10^6 cm$ where
$l_{sk}=c/w_{pe}$ is the skin depth, while the size of the knot is
less than $10$ pc \citep{hardcastle03}. The flare points and
complicated light curves \citep{hardcastle06} indicate the
disturbed effects of the irregular magnetic fields. These
small-scale random inhomogeneities give us the opportunity to
calculate the emission using perturbative DSR.  We insert Eq.(2)
to Eq.(1) and calculate numerically, we set the turbulent spectrum
$\alpha=1.45$. The electron energy distribution $dN/d\gamma
\propto\gamma ^{-s}$ is assumed as $s=3.3$. The bulk Lorenz factor
is $\Gamma=12$. The range of $k$ for turbulent spectrum
calculation can be estimated by $k_\eta/k_\nu =P_r^{1/2}$, where
the Prandtl number is $P_r\sim 10^{-5}T^4/n\sim 10^{14}$ for the
warm medium in the knot of \object{Cen A} \citep{schekochihin07}.
The final result with the comparison to the observational data is
shown in Fig. 1. Thus, we use the single gross turbulent spectrum
to reproduce the multi-band emission, with its drop-off point
properly shown in the X-ray band. From the data fitting, we find
that the relativistic electrons with $1 \leq \gamma_e \leq
10^3-10^4$ are enough for this multi-band emission, while the
turbulent magnetic field is strong, at least $10^{-3}G$, which is
larger than the equipartition value of $100\mu G$ estimated by
synchrotron radiation \citep{kataoka06}.

The radiative cooling of synchrotron emission may be one of the
reasons to explain the deeper spectrum toward high energy bands
\citep{heavens87, meisenheimer89}. The observation of \object{M87}
supports this traditional interpretation \citep{harris06}. And the
synchrotron emission by two population of electrons is needed
\citep{sambruna01} to explain the X-ray spectrum of
\object{3C273}. But for the spectrum of the knot in \object{Cen
A}, the difference of the spectral indexes between the flatter
part and the deeper part is less than 0.5 \citep{hardcastle06}.
This is contradictory to the prediction of typical synchrotron
electron cooling. For another point of view, the relatively low
number density of the knot can contribute just a small amount of
absorption, thus the strong decrease of flux in X-ray band is not
due to dust attenuation. Therefore, the drop-off point in the
spectrum might present the behavior of the turbulence.

From another side, we may directly describe the magnetic field as:
$B^2(k)\propto k^{-p}$. For this point, we avoid the detailed
treatments of any turbulence model. With the double power-law as
the form of magnetic field to calculate DSR, we select $p_1=1.4$
and $p_2=1.7$ respectively to get the result shown in the Fig. 2.
But the bulk Lorenz factor is changed from $\Gamma_1=12$ to
$\Gamma_2=2$. This result gives us an alternative clue to explain
the multi-band spectrum: the break point in the spectrum might
indicate the bulk transition state of the shock from
extra-relativistic to sub-relativistic phase.

There are some knots in other objects observed by multi-band
telescopes. Different knots have different spectral slopes and
different quantities of flux, indicating the non-uniform turbulent
mode and different acceleration processes. In this paper, we give
the example of \object{Cen A}. However, whatever the spectral
shape is, we see that the observational spectrum can be explained
by perturbative DSR theory, the emission is dominated by the
random magnetic field which could be amplified by the turbulence.
Thus, the spectral shape is uniquely determined by the random
magnetic field from radio to X-ray band.

\section{Discussion}
In this paper, we use the turbulent spectrum to amplify the random
magnetic field. We find that the spectrum shape of DSR is only
dominated by the stochastic magnetic field. The existence of this
kind of magnetic field has been confirmed by numerical simulations
\citep{haugen04a, haugen04b, schekochihin04}. Thus, the whole
multi-band radiation is produced originally from a relatively
small region, about several $pc$, with a series of physical
processes.

Furthermore, the light curves at radio, infrared and X-ray band of
the knots are more complicated. It seems that the emission is
firstly seen in the X-ray band, then followed by infrared and
radio bands \citep{hardcastle06}. We expect that the turbulent
magnetic field could have time evolution during the cascade
process with energy transfer. The final radiation spectrum may be
a composite result from the multi-structure of the turbulence
spectra and is averaged by the time evolution. Deep research of
the delicate structure in the turbulent magneto-fluid is
encouraged to explain the time-dependent features.

Three aspects are included in the whole scenario: turbulence,
magnetic field and particle acceleration. In our opinion, firstly,
the fluid background is disturbed by the relativistic
collision-less shock, the perturbative dynamos are distributed as
the turbulence spectrum; then, the initial magnetic elements are
amplified by the turbulence, shown as the random and small-scale
magnetic field; finally, the particles can be accelerated by
relativistic shocks and/or turbulent flow to produce DSR as a
first step, then continually accelerated to the higher energy part
by other mechanisms which are related to the mature magnetic
field. For simplicity, in this Letter, we assume that the electron
injection is continuous so that the spectrum does not show energy
loss by radiation.

The electron energy distribution $dN(\gamma)/d\gamma\propto
\gamma^{-s}$ has no universal index $s$ \citep{shen06}. This
suggests that particle acceleration may also have multiple
processes. There are at least two ways to accelerate electrons.
\citet{honda05} considered that the electrons are accelerated by
the interaction with the local magnetic filaments; although our
model prohibits an external magnetic field, since the
Alfv$\acute{e}$n turbulence can be self-excited by the diffusive
shocks \citep{sokolov06}, the popular Fermi and stochastic
acceleration can also be accepted in the local region. Other
models reveal that the index $s$ varies with the upstream and
downstream of the shock \citep{keshet05, baring07}. Recent
research even finds that particle acceleration is affected by the
equation of state \citep{morlino07}. Further investigation into
the relationship of turbulence, magnetic field and particle
acceleration would be expected.

\acknowledgments We thank A. Celotti and F. Yuan for the helpful
discussion. This work is financially supported by the Chinese
National Science Fund 10673028.

\clearpage

\begin{figure}
\epsscale{.80} \plotone{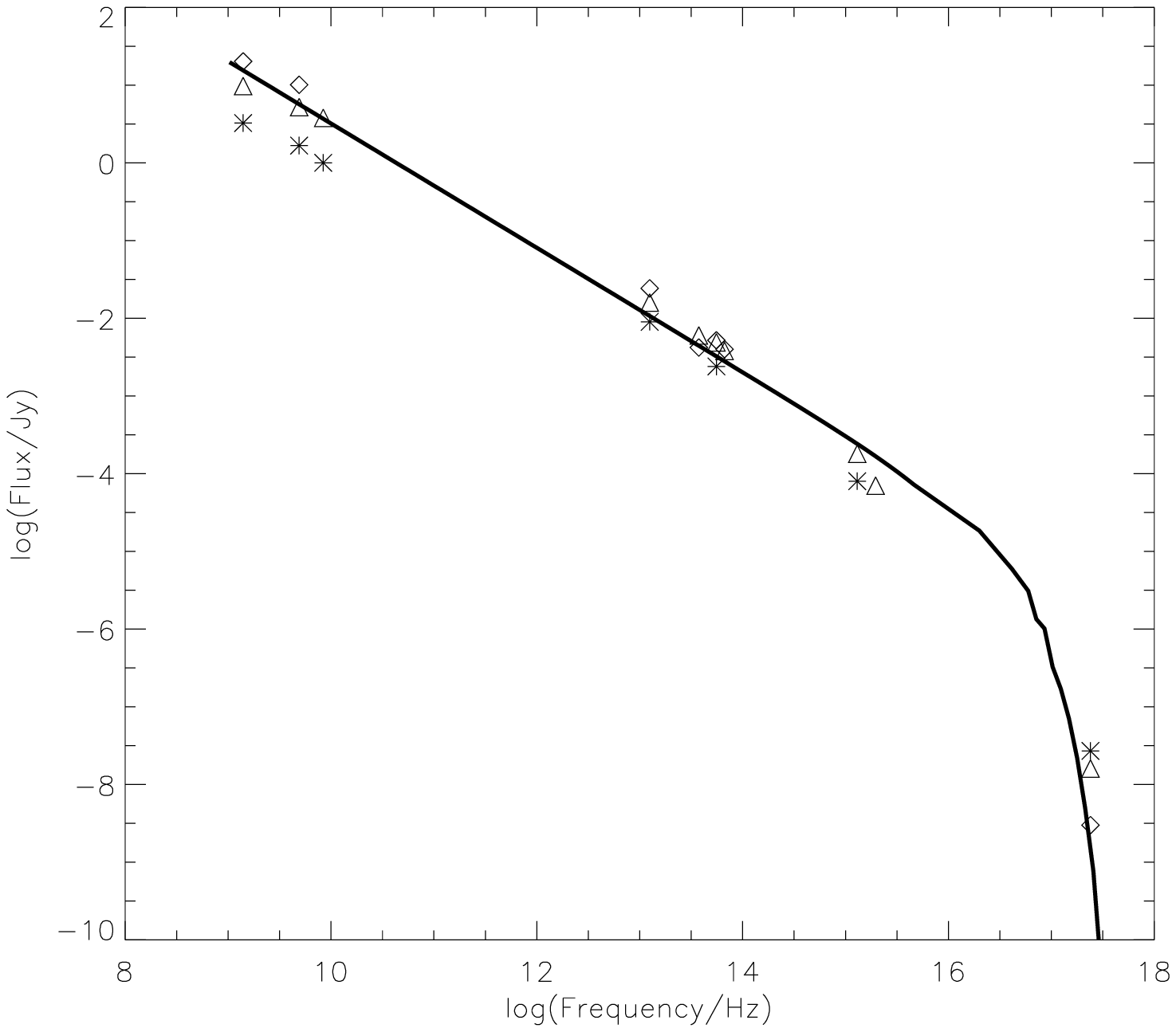} \caption{Multi-band spectrum of
knot in Cen A predicted by perturbative DSR. Magnetic field is
obtained by the Eq. (3). The observational data (inner, middle and
outer regions are symbolized by stars, triangles and diamonds
respectively) are collected from \citet{hardcastle06}. Our
prediction is denoted as the solid line. \label{fig1}}
\end{figure}

\clearpage

\begin{figure}
\epsscale{.80} \plotone{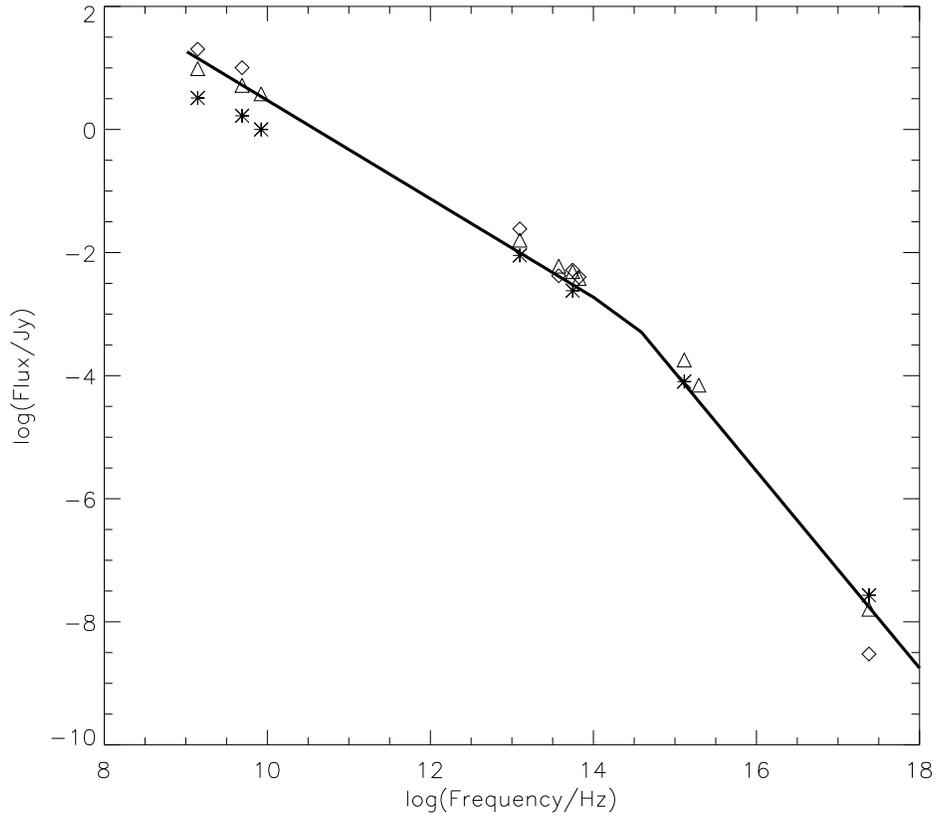} \caption{Multi-band spectrum of
knot in Cen A predicted by perturbative DSR. The solid line
denotes the radiation determined directly by the random magnetic
field of a double power-law. The observational data are shown as
same as those in the Fig.1. \label{fig2}}
\end{figure}

\clearpage

\end{document}